\documentclass[twocolumn,showpacs,amsmath,amssymb,pre]{revtex4}

\usepackage{graphicx}
\usepackage{dcolumn}
\usepackage{bm}
\begin{document}
\title{Fjords in viscous fingering: Selection of width and opening angle}
\author{Leif Ristroph$^1$}
\author{Matthew Thrasher$^1$}%
 \email{thrasher@chaos.utexas.edu}
\author{Mark B. Mineev-Weinstein$^2$}
\author{Harry L. Swinney$^1$}%
\affiliation{ $^1\textrm{Center for Nonlinear Dynamics and
Department of Physics, University of Texas at Austin, Austin, Texas 78712 USA}$ \\
$^2\textrm{Applied Physics Division, MS-P365, Los Alamos National
Laboratory, Los Alamos, New Mexico 87545 USA}$}
\date{PHYSICAL REVIEW E 74, 015201(R) (2006)}

\begin{abstract}
Our experiments on viscous fingering of air into oil contained
between closely spaced plates reveal two selection rules for the
fjords of oil that separate fingers of air. (Fjords are the building
blocks of solutions of the zero-surface-tension Laplacian growth
equation.) Experiments in rectangular and circular geometries yield
fjords with base widths $\lambda_{c}/2$, where $\lambda_{c}$ is the
most unstable wavelength from a linear stability analysis. Further,
fjords open at an angle of $8.0^{\circ}\pm 1.0^{\circ}$. These
selection rules hold for a wide range of pumping rates and fjord
lengths, widths, and directions.
\end{abstract}

 \maketitle
\begin{figure}[b]
\includegraphics[width=8cm]{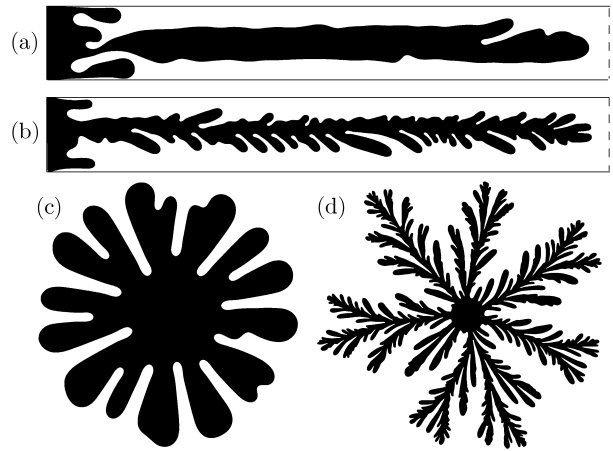}
\caption{\label{fig:ints} Viscous fingering patterns of air (black)
penetrating into oil (white) in rectangular and circular Hele-Shaw
cells for different forcing levels.  A fjord is the region of oil
between two adjacent fingers. Rectangular cell: Ca = (a)  0.0056 and
(b) 0.046; the rectangles are 25 cm wide and 190 cm long, while the
entire cell is 254 cm long. Circular cell, Ca =(c) 0.000 88 and (d)
0.27; the radial patterns are approximately 20 cm in diameter, grown
in a cell of diameter 28.8 cm. The Ca values were computed using the
average velocity of the farthest tip.}
\end{figure}

Similar growth patterns have been found for viscous fingering,
bacterial growth, flame propagation, dielectric breakdown,
electrodeposition, solidification, metal corrosion, and
diffusion-limited aggregation (DLA) \cite{Couder2000}. Analyses of
these interfacial patterns is daunting because of the broad range of
length scales, as illustrated by our experiments on fingering of air
in thin oil layers (Fig.~\ref{fig:ints}). The interface becomes
increasingly ramified with increasing forcing strength, which is
characterized by the capillary number
\begin{equation}
\label{eq:cap_num} \mathrm{Ca} = \frac{\mu V}{\sigma},
\end{equation}
where $\mu$ is the viscosity of the oil, $V$ is the local
interfacial velocity, and $\sigma$ is the surface tension of the
air-oil interface.

Most theoretical work on viscous fingering has concentrated on
fingers rather than fjords, but fjords have been considered as
building blocks of a theory of Laplacian growth in an analysis of a
two-dimensional inviscid fluid penetrating a viscous fluid (with
zero interfacial tension)~\cite{Mineev1994}. The exact solutions for
the interface are free of finite-time singularities and are linear
combinations of logarithmic terms, each term representing a single
straight fjord with parallel walls. While these analytic solutions
have helped in the understanding of the growth of complex growth
patterns, experiments indicate fjords can have curved trajectories
and nearly always have a nonzero opening angle (Fig.~\ref{fig:ints})
~\cite{Lajeunesse2000}.

Most experiments have also concentrated on fingers, but a few
experiments have examined fjords~\cite{Lajeunesse2000}. For the
circular sector (``wedge'') geometry, Thom\'e \emph{et al.}\
measured the angular gap between a wall and a divergent finger (at a
point 12 cm from the vertex) to be approximately 5$^{\circ}$ for
divergent sector angles between $20^{\circ}$ and
$50^{\circ}$~\cite{Thome1989}. Their sector decomposition method
considered finger growth in terms of wedges whose ``virtual walls''
were fjord centerlines; this approach suggests that a fjord opening
angle should be about $10^{\circ}$, i.e.\ twice the angle measured
by Thom\'e \emph{et al.}~\cite{Thome1989}. Theoretical
analysis~\cite{BenAmar,Tu1991} yielded an estimate corresponding to
a fjord opening angle of $11.7^{\circ}$~\cite{Tu1991}. Other work
has suggested that the width of a fjord relative to the width of the
splitting finger has values between $\frac{1}{8}$ and $\frac{1}{3}$
~\cite{Thome1989,Paterson1981,Pereira2004,Lajeunesse2000}.

Despite much effort, selection laws are lacking for ramified fingers
in either rectangular [Figs.~\ref{fig:ints}(a) and
~\ref{fig:ints}(b)] or circular geometries [Figs.~\ref{fig:ints}(c)
and ~\ref{fig:ints}(d)].  We find that fjords in both geometries
have a well-defined base width that is determined by the
characteristics of the interface at the fjord's birth (when a finger
splits). Further, we find that fjords open at an angle that is close
to $8^{\circ}$. The selection rules for the fjord width and opening
angle hold for a wide variation in
the ramification of the patterns. 

 \begin{figure*}[t]\includegraphics[width=18cm]{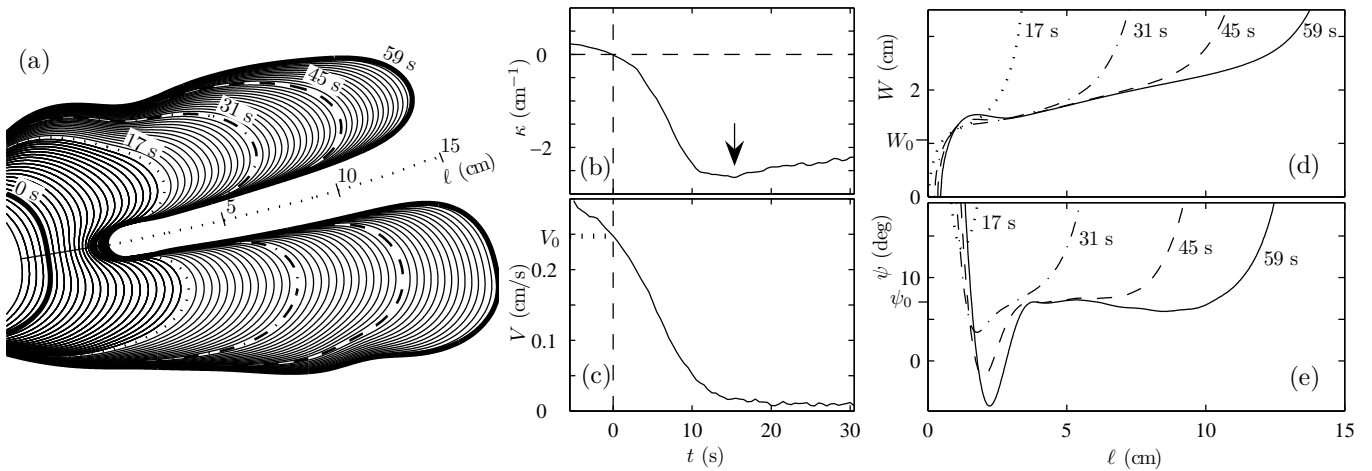}
\caption{\label{fig:exptvars} (a) Time development of a fjord grown
in the rectangular cell at Ca = 0.0076 (cf.\ Fig.\ 1 caption).
Adjacent interfaces are separated by 1 s. (b) Curvature $\kappa$ and
(c) velocity $V$ of the point on the finger ($\kappa>0$) that
evolves into a fjord ($\kappa<0$) and stops moving; $t=0$ is the
time at which $\kappa=0$. The velocity $V_0$ used in computing the
instability wavelength $\lambda_c$ [Eq.~(\ref{eq:inst_wl})]
corresponds to $\kappa=0$; here $V_0$ = 0.25 $\pm$ 0.01 cm/s. The
arrow points to the minimum of curvature in time, which defines the
fjord centerline arc length $\ell=0$. (d) Fjord width $W$ as a
function of the fjord centerline arc length $\ell$. The base width
of the fjord, $W_0$, is given by the fjord width extrapolated to
$\ell = 0$; here $W_0 = 1.08 \pm 0.06$ cm. (e) The selected opening
angle of the fjord, given by the plateau in $\psi(\ell)$, is $\psi_0
= 7.1^{\circ}\pm 0.5^{\circ}$.} %
\end{figure*}

\emph{Experimental apparatus ---} Interfacial patterns were grown in
both rectangular and circular Hele-Shaw cells (closely spaced glass
plates) filled with oil. Air was forced into the oil, creating an
unstable interface (Fig.~\ref{fig:ints}). In the rectangular
geometry, two pieces of glass (each 1.91 cm thick), separated by a
gap of 0.0508 cm, bounded a cell 25.4 cm wide by 254 cm
long~\cite{Moore2002}. The maximum variation in gap thickness at the
center of the cell (due to the imposed pressure gradient) was less
than 3\% even at the highest pump rate~\cite{Moore2002}. The cell
was filled with silicone oil (viscosity $\mu = 50.6$ mPa s and
surface tension $\sigma = 20.5$ mN/m at 24 $^\circ$C); the oil wets
the glass completely.  A uniform flow rate was achieved using a
syringe pump to withdraw oil from a reservoir at one end of the
channel; an air reservoir at atmospheric pressure was attached to
the other end. The channel was illuminated from below.  Each full
interface, as in Figs.~\ref{fig:ints}(a) and ~\ref{fig:ints}(b), was
constructed from 11 images (each $1300 \times 1030$ pixels, 0.21
mm/pixel), which were obtained using a camera and rotating mirror.
Alternatively, the camera was focused on a fixed 22-cm-long section
of the channel to obtain a time resolution of 12 frames/s [Fig.\
\ref{fig:exptvars}(a)].

In the radial geometry, air was forced into an oil-filled gap
through a hole in the center of the bottom glass plate. Each
optically polished glass plate had a diameter of 28.8 cm and a
thickness of 6.0 cm. The two plates were separated by a gap
thickness of 0.0127 cm (uniform to 1\%) \cite{Sharon2003,Praud2005}.
The gap and an annular reservoir were filled with silicone oil
(viscosity $\mu = 345$ mPa s and surface tension $\sigma = 20.9$
mN/m at 24 $^\circ$C). Interfacial patterns were grown either by
maintaining a constant pressure difference between the oil reservoir
and air or by using a syringe pump to remove oil from the buffer.
Images of resolution 0.32 mm/pixel were acquired at up to 12
frames/s.

\emph{Fjord characteristics ---} From the images we determined, as a
function of time, the interface's velocity $V$ and curvature
$\kappa$, the fjord width $W$, and the fjord opening angle $\psi$
(Fig.~\ref{fig:exptvars}). The most unstable (critical) wavelength
of a curved interface between a less viscous fluid forced into a
more viscous fluid is given by~\cite{Chuoke1959,Bataille1968}
\begin{equation}
\label{eq:inst_wl} \lambda_c = \frac{\pi}{\sqrt{\mathrm{Ca}/b^2 +
\kappa^2/12}} \thickapprox \frac{\pi b}{\sqrt{\mathrm{Ca}}}=\pi b
\sqrt{\frac{\sigma}{\mu V}},
\end{equation}
where $b$ is the cell gap thickness and $\kappa$ is the interface
curvature (not including the curvature in the small gap dimension)
at the point where the finger becomes unstable
~\cite{ParkHomsyTheory,Schwartz,Chen,footnote}. For all interfaces
that we examined, the curvature term in Eq.~(\ref{eq:inst_wl}) was
about two orders of magnitude smaller than the term involving the
capillary number Ca; hence we made the flat interface ($\kappa = 0$)
approximation in computing $\lambda_c$.

When a finger became unstable, the finger flattened and a fjord was
born. A normal projection algorithm was used to track the fjord
trajectory back in time to earlier interfaces [solid, nearly
straight line on the left in Fig.~\ref{fig:exptvars}(a)]. (The fjord
location for an earlier interface was identified by projecting a ray
normally from the later interface until it intersected the earlier
one.) The time evolution of the local curvature $\kappa$ then
revealed the birth of the fjord, where $\kappa=0$
[Fig.~\ref{fig:exptvars}(b)].  The velocity $V_0$ at this point
[Fig.~\ref{fig:exptvars}(c)] was used in Eq.~(\ref{eq:inst_wl}) to
calculate $\lambda_c$.  The uncertainty in $\lambda_c$ is 9\%; film
wetting corrections~\cite{Schwartz,ParkHomsyTheory} and the velocity
measurements are the main sources of uncertainty.

\begin{figure}[t]
\includegraphics[width=8cm]{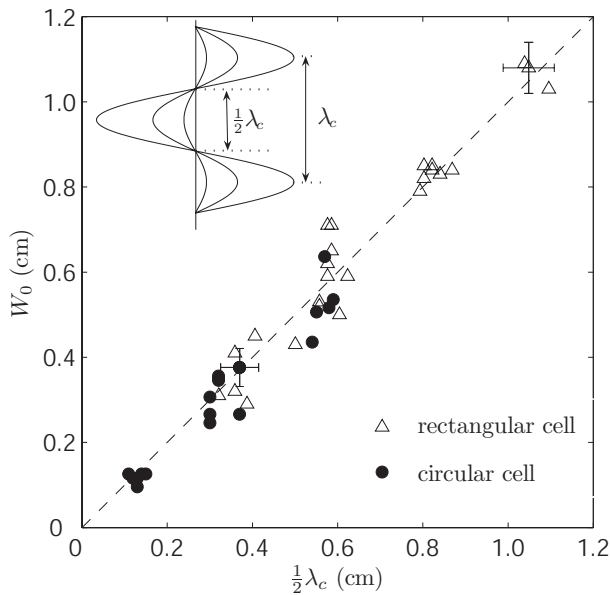}
\caption{\label{fig:W_0} The measured fjord base width $W_0$
[Fig.~\ref{fig:exptvars}(d)] versus the predicted width,
$\lambda_c/2$, where $\lambda_c$ is the instability wavelength,
Eq.~(\ref{eq:inst_wl}). The inset illustrates how an initial
infinitesimal perturbation of a front sets the wavelength for a
large-amplitude fjord.  Here, the front is moving to the right and
is observed in a comoving frame. The lengths of the error bars
are the rms deviations from the line.}
\end{figure}

Once a fjord developed, we determined its width $W$ as a function of
the arc length $\ell$ of its centerline, as illustrated in
Fig.~\ref{fig:exptvars}(d).  The fjord centerline was found by
beginning with a point on the side of the fjord; then the closest
point on the opposite side was found such that the interior angles
between the two local tangents and the line connecting the two
points were the same.  The midpoint of the connecting line gave a
point on the fjord centerline. Repeating this procedure for each
point on the side of the fjord constructed the centerline. The fjord
base point ($\ell=0$) was taken to correspond to the minimum of
curvature along the fjord's trajectory [Fig.~\ref{fig:exptvars}(b)].
As surface tension relaxed the interface, a fjord's base moved
forward. The distance a base moved after $t=0$ was corrected by
adding the same distance to the arc length
[Fig.~\ref{fig:exptvars}(d)].

Curves $W(\ell)$ obtained for different fjords exhibited the same
structure: $W(\ell)$ initially grew quickly with increasing $\ell$
due to the rounded nature of the fjord base, but soon $W(\ell)$
increased linearly with $\ell$. The fjord base width $W_0$ was
measured as follows: for each $\ell$ the local tangent of $W(\ell)$
was linearly extrapolated back to $\ell=0$, a histogram of
extrapolated base widths was made, and the histogram's peak gave
$W_0$.  For an individual fjord measured over a wide range of times,
this procedure yielded a fjord base width with a typical uncertainty
of 6\%. Confinement by neighboring fjords caused a slightly
increasing extrapolated fjord base width with increasing values of
$\ell$.

The observed linear increase in $W(\ell)$ suggested an almost
constant opening angle: $\psi =\arctan(\frac{dW}{d\ell})$. The angle
$\psi(\ell)$ had the same qualitative evolution for all fjords: a
fjord began with an opening angle of 180$^\circ$, followed by a
surface-tension-induced dip, followed by a peak and a plateau
(usually with a slightly negative slope), and finally a rapid
increase at the fjord's end [Fig.~\ref{fig:exptvars}(e)]. We defined
the opening angle $\psi_0$ to be the peak of the histogram of $\psi$
values generated from plots of $\psi(\ell)$. The angle selected was
always about $8^\circ$; the uncertainty in the opening angle for a
single fjord was 0.5$^{\circ}$.

\emph{Fjord base width ---} We predict that the base width of a
fjord is determined at the onset of the instability of the moving
interface, as illustrated by the inset of Fig.~\ref{fig:W_0}. The
critical wavelength $\lambda_c$ given by a linear stability analysis
of the infinitesimally perturbed flat front, Eq.~(\ref{eq:inst_wl}),
determines the distance between adjacent fingers when they are born,
and the base width of the macroscopic fjord that develops between
the two emergent fingers is $W_0 = \frac{1}{2} \lambda_c$.  A
comparison between measurements and the predicted widths requires a
value for the velocity $V$ in the expression for the capillary
number Ca. Lajeunesse and Couder suggest that a perturbation on a
finger is the ``precursor'' of tip splitting and confirm that the
perturbation's wavelength is ``of the order of 1.5
$\lambda_c$~\cite{Lajeunesse2000}.'' Our approach tests this idea at
the instability's onset, making the assumption that $V=V_0$ is
measured at the point on the fjord's trajectory corresponding to
$\kappa=0$ [Figs.~\ref{fig:exptvars}(b) and ~\ref{fig:exptvars}(c)].

Measurements of the fjord base widths $W_0$ in both the circular and
rectangular geometries agree well with the predicted widths, as
Fig.~\ref{fig:W_0} illustrates.  The fjord widths varied by an order
of magnitude as the control conditions were varied widely (pump
rates were 0.17 - 2.00 cm$^3$/s in the rectangular cell and 0.000 83
- 0.033 cm$^3$/s in the circular cell, and in the latter geometry,
measurements were also made for pressure differences of 0.10 and
0.25 atm). The local forcing strength at the birth of a fjord is
affected by nearby fjords, walls, and details of the interface, so a
given global pumping rate can yield fjords with different
properties. The uncertainties in $\lambda_c$ and $W_0$ are 9\% and
6\%, respectively. The observed scatter in Fig.\ 3 is consistent
with these estimated errors.

\begin{figure}[t]
\includegraphics[width=8cm]{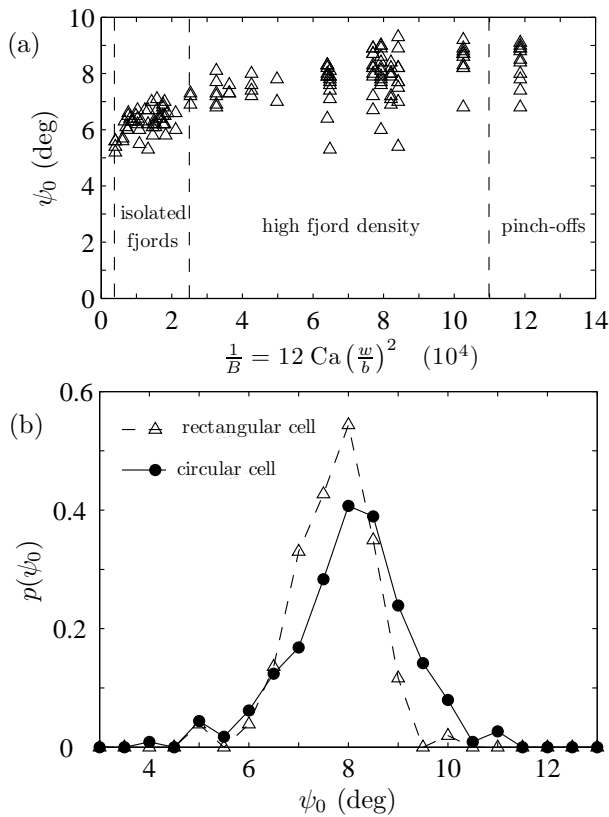}
\caption{\label{fig:psiB} (a) The fjord opening angle $\psi_0$ in
the rectangular cell increases slightly with forcing parameter (the
modified capillary number $1/B$). (b) Probability distributions
$p(\psi_0)$ measured for the opening angle of all fjords observed in
the circular cell ($\bullet$) and of fjords in the
high-fjord-density regime ($1/B >$ 25 000) of the rectangular cell
($\triangle$). The data for the rectangular and circular cells
yield, respectively, $\psi_0 = 7.9^{\circ}\pm0.8^{\circ}$ and
$\psi_0 = 8.2^{\circ}\pm1.1^{\circ}$, where the uncertainties are
one standard deviation.}
\end{figure}

\emph{Fjord opening angle ---} The opening angle $\psi_0$ measured
in the rectangular geometry is shown in Fig.~\ref{fig:psiB}(a) as a
function of the appropriate forcing parameter for this geometry, the
modified capillary number $1/B = 12 Ca (w/b)^2$ (where $w$ is the
channel width)~\cite{Park1985,Moore2002}, which was computed using
the average velocity of the farthest finger tip. Isolated fjords
occur in the regime just beyond the onset of tip splitting, which
occurs for $1/B \approx 4000~$ \cite{Park1985,footnote2}.

For $1/B >$ 25 000, the pattern develops a high density of fjords.
Despite the wide range of experimental parameters in this regime,
the measured angles are surprisingly similar. The results for the
distribution of opening angles in the circular cell are
indistinguishable from those for ramified patterns in the
rectangular cell: the probability distribution function $p(\psi_0)$
has the same form with essentially the same mean 8$^{\circ}$ and
same standard deviation 1$^{\circ}$ [Fig.~\ref{fig:psiB}(b)]. We
suggest that the near constancy of the opening angle is fundamental
to scale invariance and fractal dimension. Our result is consistent
both with efforts to relate the multifractal dimensions of a fully
developed fingering structure~\cite{Mathieson} with the fjords'
opening angle~\cite{Sarkar} and with a recent
observation~\cite{Praud2005} of an invariant unscreened angle
distribution in a fractal grown in a radial geometry.  While
intriguing, the connection between these invariant geometrical
characteristics remains obscure.

\emph{Conclusions ---} We have presented two selection rules for the
geometric form of fjords in rectangular and circular cells. The base
width selected by the fjords has been found to be well described by
$W_0 = \frac{1}{2} \lambda_c$, where $\lambda_c$ is the local
critical wavelength. This selection rule provides insight into the
most apparent difference in viscous fingering patterns: highly
forced patterns are composed of thinner fjords.  The rule is based
on universal features of Laplacian growth and does not invoke any
specific properties of viscous fingering in a Hele-Shaw cell.
Therefore, we conjecture that this rule should apply not only to
viscous fingering, but also to other isotropic Laplacian growth
problems.

The second selection rule is obtained from measurements of the
opening angle as a function of arc length along the fjord
centerline: for a broad range of experimental parameters and for
both circular and rectangular geometries, the selected opening angle
$\psi_0$ was always near 8$^{\circ}$; we have no explanation for
this result. The two selection rules hold for a wide range of fjord
widths, lengths, and degrees of bending, and for a wide range of
angles of fjord directions with respect to the rectangular cell axis
or with respect to radial lines in the circular cell.

We thank W. D. McCormick, M. Moore, O. Praud, and J. Swift for
helpful discussions.  This work was supported by an Office of Naval
Research Quantum Optics Initiative Grant and by a Los Alamos
National Laboratory LDRD Grant.

\bibliography{biblio}

\begin{thebibliography}{21}
\bibitem[{foo({\natexlab{b}})}]{footnote2}
\bibinfo{note}{The isolated fjords have a selected opening angle $\psi_0 =
  6.3^{\circ}\pm0.4^{\circ}$, which is lower than the angle measured in fjords
  grown with $1/B >$ 25 000. These fjords are strongly confined by the walls.}
\end{thebibliography}
\end{document}